\pdfoutput=1 

\documentclass[aps,prl,twocolumn,showpacs,preprintnumbers,letterpaper,superscriptaddress,amsmath,amssymb]{revtex4-1}

\usepackage[utf8]{inputenc} 
\usepackage{graphicx} 
\usepackage{dcolumn} 
\usepackage{bm} 
\usepackage{todonotes}

\newcommand{\mbold}[1]{\mbox{\boldmath${#1}$}}
\renewcommand{\vec}[1]{\mathbf{{#1}}}
\newcommand{\pvec}[1]{\mathbf{{#1}}_\parallel }

\newcommand{\nn}{\nonumber}
\newcommand{\imu}{\mathrm{i}}
\newcommand{\dint}{\mathrm{d}}
\newcommand{\inc}{\ensuremath{(i)}}
\newcommand{\sca}{\ensuremath{(s)}}

\renewcommand{\Re}{\mathrm{Re}}
\renewcommand{\Im}{\mathrm{Im}}

\begin{document}


\title{Calculation of all elements of the Mueller matrix for scattering
       of light from a two-dimensional randomly rough metal surface}

\author{P. A. Letnes}
\email{Paul.Anton.Letnes@gmail.com}
\affiliation{Department of Physics, Norwegian University of
 Science and Technology, NO-7491 Trondheim, Norway}

\author{A. A. Maradudin}
\email{aamaradu@uci.edu}
\affiliation{Department of Physics and Astronomy, and
 Institute for Surface and Interface Science, University of
 California, Irvine, CA 92697 USA}

\author{T. Nordam}
\email{Tor.Nordam@ntnu.no}
\affiliation{Department of Physics, Norwegian University of
 Science and Technology, NO-7491 Trondheim, Norway}

\author{I. Simonsen}
\email{Ingve.Simonsen@ntnu.no}
\affiliation{Department of Physics, Norwegian University of
 Science and Technology, NO-7491 Trondheim, Norway}

\begin{abstract}
We calculate all the elements of the Mueller matrix, which contains
all the polarization properties of light scattered from a
two-dimensional randomly rough lossy metal surface.  The
calculations are carried out for arbitrary angles of incidence by
the use of nonperturbative numerical solutions of the reduced
Rayleigh equations for the scattering of p- and s-polarized light
from a two-dimensional rough penetrable surface. The ability to model polarization effects in light scattering from surfaces enables better interpretation of experimental data and allows for the design of surfaces which possess useful polarization effects.
\end{abstract}

\maketitle


When light undergoes scattering from a surface, the scattered light carries a great deal of information about the statistical properties of the surface in its polarization. Even when the structures in question are sub-wavelength and beyond the imaging limit, polarized optical scattering can be employed to detect and distinguish between material inhomogeneities, particles, or even buried defects and the roughness of both interfaces of thin films~\cite{springerlink:10.1007/978-0-387-35659-4_10}. These techniques are already in use in the semiconductor industry, and such techniques could become important for surface characterization of photovoltaic materials and nanomaterials~\cite{Drezet:08,Kullock:08}. Many biological materials are optically active, meaning that polarimetric measurements may be applied for characterization of biological or hybrid~\cite{Schmidt20112645} materials and even for the search for extra-terrestrial life~\cite{Martin20102444}. Surface patterning has also been proven as a method for creating optical components with interesting polarization properties~\cite{Ghadyani:11}.

To extract this information from experimental data, one has to be able to model polarization effects~\cite{Ellis:02}. The ability to calculate the polarization of the scattered light also opens the door to the possibility of designing surfaces that produce specified polarization properties of the light scattered from them.

All the information about the polarization properties of light
scattered from two-dimensional surfaces is contained in the
Mueller matrix~\cite{Mueller,bickel:468,chipman}.  Yet, very few calculations of the
elements of this matrix for a two-dimensional randomly rough surface
have to date been carried out by any computational approach, largely
because calculations of the scattering of light from such surfaces are
still difficult to carry out~\cite{PhysRevA.81.013806,PhysRevLett.104.223904,nordam-letnes-simonsen,Simonsen:2011aa}.
An exception~\cite{doi:10.1080/13616679809409826} is a calculation of the Mueller matrix for two-dimensional randomly rough perfectly conducting and metallic
surfaces characterized by a surface profile function that is a
stationary, zero-mean, isotropic, Gaussian random process, defined by
a Gaussian surface height autocorrelation function.  These
calculations were carried out by a ray-tracing approach on the
assumption that the surface was illuminated at normal incidence.  In
this work it was also shown that due to the assumptions of normal
incidence and the isotropy of the surface statistics, the elements of
the corresponding Mueller matrix possess certain symmetry properties.
Subsequently Zhang and Bahar~\cite{774161} carried out an
approximate analytic calculation of the elements of the Mueller matrix
for the scattering of light from two-dimensional randomly rough
dielectric surfaces coated uniformly with  a different
dielectric material.

In this Letter we report the first step toward realizing the possibilities mentioned in the opening paragraphs.  We present an approach to calculating, for arbitrary angles of incidence, all the elements of the Mueller matrix for the scattering of light from a two-dimensional randomly rough metal surface.  It is based on nonperturbative numerical solutions of the reduced Rayleigh equations for the scattering of p- and s-polarized light from a two-dimensional rough penetrable surface \cite{nordam-letnes-simonsen,Brown1984381}.

The system we study consists of vacuum in the region $x_3 >
\zeta(\pvec{x})$, where $\pvec{x} = (x_1, x_2,0)$, and a metal whose
dielectric function is $\varepsilon(\omega)$ in the region $x_3 <
\zeta(\pvec{x})$.  The surface profile function $\zeta(\pvec{x})$ is
assumed to be a single-valued function of $\pvec{x}$ that is
differentiable with respect to $x_1$ and $x_2$, and constitutes a
stationary, zero-mean, isotropic, Gaussian random process defined by
$\left< \zeta(\pvec{x}) \zeta(\pvec{x}') \right> = \delta^2
W \left( \left| \pvec{x} - \pvec{x}' \right| \right)$.  The angle brackets here and in all that
follows denote an average over the ensemble of realizations of the
surface profile function, and $\delta =\left< \zeta^2(\pvec{x}
  )\right>^{1 / 2}$ is the rms height of the surface.  Each
realization of the surface profile function was generated numerically
by the filtering method used in Refs.~\cite{Simonsen:2011aa,Maradudin1990255}.

We begin by writing the electric field in the vacuum region $x_3 >
\zeta(\pvec{x})$ as the sum of an incident and a scattered field,
$\vec{E}(\vec{x}, t) = \left[\vec{E}^{\inc}(\vec{x} |\omega
)+\vec{E}^{\sca}(\vec{x} |\omega )\right] \exp (-\imu\omega t)$, where
\begin{widetext}
\begin{subequations}
\begin{align*}
    \vec{E}^{\inc}(\vec{x} |\omega ) &=
    \left[   {\mathcal E}^{\inc}_p(\pvec{k} )  \vec{\hat{e}}^{\inc}_p(\pvec{k} )
        + {\mathcal E}^{\inc}_s(\pvec{k} )  \vec{\hat{e}}^{\inc}_s(\pvec{k} )
    \right]
    \exp\left[
        \imu \pvec{k}\cdot\pvec{x} -\imu \alpha_0(k_\parallel) x_3
    \right],
    \\
    \vec{E}^{\sca}(\vec{x} |\omega )
     &= \int\!\! \frac{\dint^2q_\parallel}{(2\pi )^2}\:
      \left[
            {\mathcal E}^{\sca}_p(\pvec{q})  \vec{\hat{e}}^{\sca}_p(\pvec{q} )
        +   {\mathcal E}^{\sca}_s(\pvec{q} ) \vec{\hat{e}}^{\sca}_s(\pvec{q} )
       \right]
    \exp [\imu \pvec{q} \cdot \pvec{x} +\imu \alpha_0(q_\parallel) x_3 ].
\end{align*}
\end{subequations}
\end{widetext}
Here $\pvec{k} = (k_1,k_2,0)$, the unit polarization vectors are
$\vec{\hat{e}}^{\inc}_p(\pvec{k}) =
    (c/\omega ) \left[ \alpha_0(k_\parallel)\vec{\hat{k}}_\parallel +k_\parallel \vec{\hat{x}}_3 \right]$,
$\vec{\hat{e}}^{\inc}_s(\pvec{k} ) = \vec{\hat{k}}_\parallel \times \vec{\hat{x}}_3$,
$\vec{\hat{e}}^{\sca}_p(\pvec{q} ) =
 (c/\omega ) \left[ -\alpha_0(q_\parallel) \vec{\hat{q}}_\parallel + q_\parallel\vec{\hat{x}}_3 \right]$,
$\vec{\hat{e}}^{\sca}_s(\pvec{q}) = \vec{\hat{q}}_\parallel \times \vec{\hat{x}}_3$,
while $\alpha_0(q_\parallel) = \left[ (\omega /c)^2-q_\parallel^2 \right]^{1 / 2}$,
with $\Re\,\alpha_0(q_\parallel) > 0$, $\Im\,\alpha_0(q_\parallel) > 0$.  A
caret over a vector indicates that it is a unit vector.  In terms of
the polar and azimuthal angles of incidence $(\theta_0,\phi_0)$ and
scattering $(\theta_s,\phi_s)$, the vectors $\pvec{k}$ and $\pvec{q}$
are given by $\pvec{k} = (\omega /c)\sin\theta_0
(\cos\phi_0,\sin\phi_0,0)$ and $\pvec{q} = (\omega /c)\sin\theta_s
(\cos\phi_s ,\sin\phi_s ,0)$.

A linear relation exists between the amplitudes ${\mathcal E}^{(s)}_{\alpha}(\pvec{q} )$
and ${\mathcal E}^{(i)}_{\beta}(\pvec{k} )$, which we write in the form $(\alpha =
p,s$, $\beta = p,s)$
\begin{align*}
    {\mathcal E}^{(s)}_{\alpha}(\pvec{q} ) &=
    \sum_{\beta}R_{\alpha\beta}(\pvec{q} |\pvec{k} )
        {\mathcal E}^{(i)}_{\beta}(\pvec{k} ).
\end{align*}
It was shown by Celli and his colleagues~\cite{Brown1984381} that the scattering
amplitudes $R_{\alpha\beta}(\pvec{q} |\pvec{k} )$ satisfy the matrix
integral equation (the reduced Rayleigh equation)
\begin{align}
    \int & \frac{\dint^2 q_\parallel}{(2\pi )^2}\;
    \frac{
        I\left(
            \alpha (p_\parallel ) - \alpha_0(q_\parallel) |\pvec{p} - \pvec{q}
        \right)
    }{
        \alpha(p_\parallel) - \alpha_0(q_\parallel)
    }
    \mbold{\mathcal N}_{+}(\pvec{p} | \pvec{q} )\vec{R} (\pvec{q} |\pvec{k} )
    \nn\\
    &\quad = -\frac{
    I\left(
        \alpha(p_\parallel) + \alpha_0(k_\parallel)|\pvec{p} - \pvec{k}
    \right)
    }{
    \alpha(p_\parallel) + \alpha_0(k_\parallel)
    }
    \mbold{\mathcal N}_{-}(\pvec{p} |\pvec{k}),
    \label{eq:3}
\end{align}
with $R_{pp}$ and $R_{ps}$ forming the first row of the matrix $\vec{R}$, where
\begin{align}
    I \left( \gamma |\pvec{Q} \right) = \int \dint^2x_\parallel\;
        \exp [-\imu\gamma\zeta(\pvec{x}) ]
            \exp (-\imu\pvec{Q} \cdot\pvec{x} ),
    \label{eq:4}
\end{align}
and $\alpha(p_\parallel) = \left[\varepsilon(\omega)(\omega /c)^2
-p_\parallel^2 \right]^{1/2}$, with $\Re\,\alpha(p_\parallel) > 0$,
$\Im\,\alpha(p_\parallel) > 0$.
The matrices $\mbold{\mathcal N}_{\pm }\left( \pvec{p} |\pvec{q} \right)$ are given by
\begin{align*}
    \mbold{\mathcal N}_{\pm} & (\pvec{p} |\pvec{q} ) =
    \nn \\
    & \begin{pmatrix}
        p_\parallel q_\parallel \pm \alpha(p_\parallel) \vec{\hat{p}}_\parallel \cdot \vec{\hat{q}}_\parallel \alpha_0(q_\parallel)
        &
        -\frac{\omega}{c} \alpha(p_\parallel) (\vec{\hat{p}}_\parallel \times \vec{\hat{q}}_\parallel)_3
        \\
        \pm \frac{\omega}{c} (\vec{\hat{p}}_\parallel \times \vec{\hat{q}}_\parallel )_3 \alpha_0(q_\parallel)
        & \frac{\omega^2}{c^2} \vec{\hat{p}}_\parallel \cdot \vec{\hat{q}}_\parallel
    \end{pmatrix}.
\end{align*}
These equations were solved by the method described in detail in~\cite{nordam-letnes-simonsen}.
First, a realization of the surface profile function on a grid of $N^2_x$ points within a square region of the $x_1x_2$ plane of edge $L$. In evaluating the $\pvec{q}$-integral in Eq.~(\ref{eq:3}) the infinite limits of integration were replaced by finite ones: $\left| \pvec{q} \right| < Q/2$, and the integral was carried out by a two-dimensional version of the extended midpoint rule~\cite{numerical-recipes} using a grid in the $q_1 q_2$ plane that is determined by the Nyquist sampling theorem and the properties of the discrete Fourier transform.
The function $I(\gamma |\pvec{Q} )$ was evaluated by expanding the integrand in Eq.~(\ref{eq:4}) in powers of $\zeta(\pvec{x})$ and calculating the Fourier transform of $\zeta^n(\pvec{x} )$ by the Fast Fourier Transform. The resulting equations were solved by LU factorization.

The scattering amplitudes $R_{\alpha\beta}(\pvec{q} |\pvec{k} )$ play
a central role in the calculation of the elements of the Mueller
matrix.  In terms of these amplitudes the elements of the Mueller
matrix, $\vec{M}$,  are~\footnote{A. A. Maradudin (unpublished)}
\begin{subequations}
\label{eq:6}
\begin{align*}
  M_{11} &= C\left(|R_{\mathrm{pp}}|^2 + |R_{\mathrm{sp}}|^2 + |R_{\mathrm{ps}}|^2 + |R_{\mathrm{ss}}|^2\right) \\
  M_{12} &= C\left(|R_{\mathrm{pp}}|^2 + |R_{\mathrm{sp}}|^2-|R_{\mathrm{ps}}|^2-|R_{\mathrm{ss}}|^2\right) \\
  M_{13} &= C\left(R_{\mathrm{pp}}R_{\mathrm{ps}}^* + R_{\mathrm{sp}}R_{\mathrm{ss}}^* + R_{\mathrm{ps}}R_{\mathrm{pp}}^* + R_{\mathrm{ss}}R_{\mathrm{sp}}^*\right) \\
  M_{14} &= \imu C \left(R_{\mathrm{pp}}R^*_{\mathrm{ps}} + R_{\mathrm{sp}}R^*_{\mathrm{ss}} - R_{\mathrm{ps}}R^*_{\mathrm{pp}}-R_{\mathrm{ss}}R^*_{\mathrm{sp}}\right) \\
  M_{21} &= C\left(|R_{\mathrm{pp}}|^2-|R_{\mathrm{sp}}|^2 + |R_{\mathrm{ps}}|^2 - |R_{\mathrm{ss}}|^2\right) \\
  M_{22} &= C\left(|R_{\mathrm{pp}}|^2-|R_{\mathrm{sp}}|^2 - |R_{\mathrm{ps}}|^2 +|R_{\mathrm{ss}}|^2\right) \\
  M_{23} &= C\left(R_{\mathrm{pp}}R^*_{\mathrm{ps}} - R_{\mathrm{sp}}R^*_{\mathrm{ss}} + R_{\mathrm{ps}}R^*_{\mathrm{pp}}-R_{\mathrm{ss}}R^*_{\mathrm{sp}}\right) \\
  M_{24} &= \imu C\left(R_{\mathrm{pp}}R^*_{\mathrm{ps}}-R_{\mathrm{sp}}R^*_{\mathrm{ss}} - R_{\mathrm{ps}}R^*_{\mathrm{pp}}+R_{\mathrm{ss}}R^*_{\mathrm{sp}}\right) \\
  M_{31} &= C\left(R_{\mathrm{pp}}R^*_{\mathrm{sp}}+R_{\mathrm{sp}}R^*_{\mathrm{pp}}+R_{\mathrm{ps}}R^*_{\mathrm{ss}}+R_{\mathrm{ss}}R^*_{\mathrm{ps}}\right) \\
  M_{32} &= C\left(R_{\mathrm{pp}}R^*_{\mathrm{sp}} + R_{\mathrm{sp}}R^*_{\mathrm{pp}}-R_{\mathrm{ps}}R^*_{\mathrm{ss}}-R_{\mathrm{ss}}R^*_{\mathrm{ps}}\right) \\
  M_{33} &= C\left(R_{\mathrm{pp}}R^*_{\mathrm{ss}}+R_{\mathrm{sp}}R^*_{\mathrm{ps}}+R_{\mathrm{ps}}R^*_{\mathrm{sp}}+R_{\mathrm{ss}}R^*_{\mathrm{pp}}\right) \\
  M_{34} &= \imu C \left(R_{\mathrm{pp}}R^*_{\mathrm{ss}}+R_{\mathrm{sp}}R^*_{\mathrm{ps}} - R_{\mathrm{ps}}R^*_{\mathrm{sp}}-R_{\mathrm{ss}}R^*_{\mathrm{pp}}\right) \\
  M_{41} &= -\imu C \left(R_{\mathrm{pp}}R^*_{\mathrm{sp}} - R_{\mathrm{sp}}R^*_{\mathrm{pp}}+R_{\mathrm{ps}}R^*_{\mathrm{ss}}-R_{\mathrm{ss}}R^*_{\mathrm{ps}}\right) \\
  M_{42} &= -\imu C \left(R_{\mathrm{pp}}R^*_{\mathrm{sp}}-R_{\mathrm{sp}}R^*_{\mathrm{pp}}-R_{\mathrm{ps}}R^*_{\mathrm{ss}}+R_{\mathrm{ss}}R^*_{\mathrm{ps}}\right) \qquad \\
  M_{43} &= -\imu C \left(R_{\mathrm{pp}}R^*_{\mathrm{ss}}-R_{\mathrm{sp}}R^*_{\mathrm{ps}}-R_{\mathrm{ps}}R^*_{\mathrm{sp}}-R_{\mathrm{ss}}R^*_{\mathrm{pp}}\right) \\
  M_{44} &= C\left(R_{\mathrm{pp}}R^*_{\mathrm{ss}}-R_{\mathrm{sp}}R^*_{\mathrm{ps}}-R_{\mathrm{ps}}R^*_{\mathrm{sp}}+R_{\mathrm{ss}}R^*_{\mathrm{pp}} \right)
\end{align*}
where
\begin{align*}
  C &= \frac{1}{2L^2} \left( \frac{\omega}{2\pi c}\right)^2 \frac{\cos^2\theta_s}{\cos\theta_0},
 \label{eq:7}
\end{align*}
\end{subequations}
and $L^2$ is the area of the plane $x_3 = 0$ covered by the rough surface.

As we are concerned with scattering from a randomly rough surface, it
is the average, $\left< \vec{M} \right>$, of the Mueller matrix over
the ensemble of realizations of the surface profile function that we
seek.  In evaluating an average of the form
$\big\langle R_{\alpha\beta}R^*_{\gamma\delta}\big\rangle$ we can write
$R_{\alpha\beta}$ as the sum of its mean value and its fluctuation
about the mean, $R_{\alpha\beta} = \big\langle R_{\alpha\beta}\big\rangle +
\left( R_{\alpha\beta}-\big\langle R_{\alpha\beta}\big\rangle \right)$.  We then obtain the result $\big\langle R_{\alpha\beta}R^*_{\gamma\delta}\big\rangle =
\big\langle R_{\alpha\beta}\big\rangle\big\langle R^*_{\gamma\delta}\big\rangle +
\left(\big\langle R_{\alpha\beta}R^*_{\gamma\delta}\big\rangle -\big\langle
  R_{\alpha\beta}\big\rangle\big\langle R^*_{\gamma\delta}\big\rangle \right)$.
The first term on the right hand side of this equation arises in the
contribution to an element of the ensemble averaged Mueller matrix
from the light scattered coherently (specularly); the second term
arises in the contribution to that ensemble averaged matrix element
from the light scattered incoherently (diffusely). It is the latter
contribution, $\left<\vec{M}\right>_{\text{incoh}}$, that we calculate.

\begin{figure*}[th]
 \centering
 \includegraphics*[width=1.65\columnwidth]{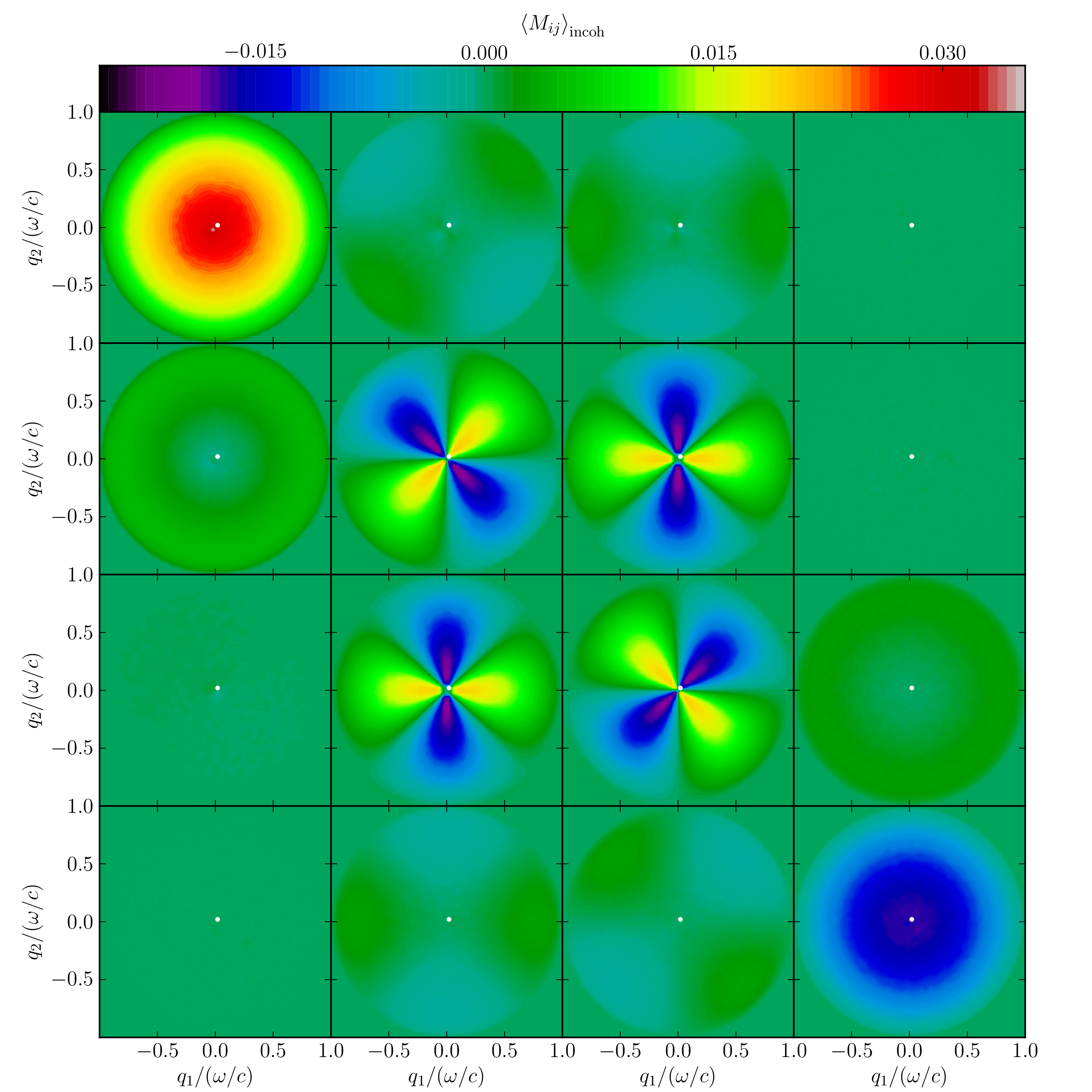}
 \caption{(Color online) Color-level plots of the
   contribution to the Mueller matrix elements from the light
   scattered incoherently
   as functions of $q_1$ and $q_2$ for angles of incidence
   $(\theta_0,\phi_0)=(2^\circ, 45^\circ)$. An ensemble consisting of
   $N_p=10\,000$ 
   surface realizations was used in obtaining these results.  The
   elements, $\left<M_{ij}\right>_{\text{incoh}}$ ($i,j=1,2,3,4$), are
   organized as a matrix with $\left<M_{11}\right>_{\text{incoh}}$ in
   the top left corner; $\left<M_{12}\right>_{\text{incoh}}$ top row
   and second column, \textit{etc}. The white spots indicate the specular
   direction in reflection.  }
 \label{Fig:1}
\end{figure*}

We have calculated in this way the $16$ elements of the Mueller matrix
when light of wavelength $\lambda = 457.9$~nm is incident on a
two-dimensional randomly rough silver surface whose dielectric
function at this wavelength is $\varepsilon(\omega) =
-7.5+\imu 0.24$~\cite{Johnson:1972kl}.
The roughness of the surface is defined by a surface height
autocorrelation function $W \left( \left|\pvec{x} \right| \right) = \exp (-x_\parallel^2
/a^2)$, where $a = \lambda /4$ and the rms height $\delta = \lambda
/40$.  For the numerical parameters we used $L=25\lambda$ and
$N_x=319$ which implies that $Q=6.4(\omega/c)$.  For these
parameters, and when the metal is assumed to be non-absorbing
[$\Im\,\varepsilon(\omega)\equiv 0$], our simulation approach
conserved energy within a margin of $1$\% or better.  Moreover, the
calculated Mueller matrices were found to be physically realizable and
therefore self-consistent by the method of
Ref.~\cite{cloude1989}.

The results presented in Fig.~\ref{Fig:1} were obtained for angles of incidence $(\theta_0,\phi_0)=(2^\circ, 45^\circ)$, \textit{i.e.} for (essentially) normal incidence. The first thing to notice from Fig.~\ref{Fig:1} is that the individual matrix elements possess the symmetry properties predicted by Bruce~\cite{doi:10.1080/13616679809409826}. The elements of the first and last column are circularly symmetric; each element of the second and third columns is invariant under a combined $90^\circ$ rotation about the origin and a change of sign; and the elements of the second column are $45^\circ$ rotations of the elements of the third column in the same row~\footnote{It should be noted that the scattering amplitudes $f_{sp}$ and $f_{ps}$ in Bruce's Eq.~(1) corresponds to our $R_{ps}$ and $R_{sp}$, respectively.}.  Note that the elements $\left<M_{31}\right>_{\text{incoh}}$, $\left<M_{41}\right>_{\text{incoh}}$, $\left<M_{14}\right>_{\text{incoh}}$, and $\left<M_{24}\right>_{\text{incoh}}$ are zero to the precision used in this calculation. However, simulations indicate that this does not hold for anisotropic surfaces.

\begin{figure}[htb]
    \centering
    \includegraphics{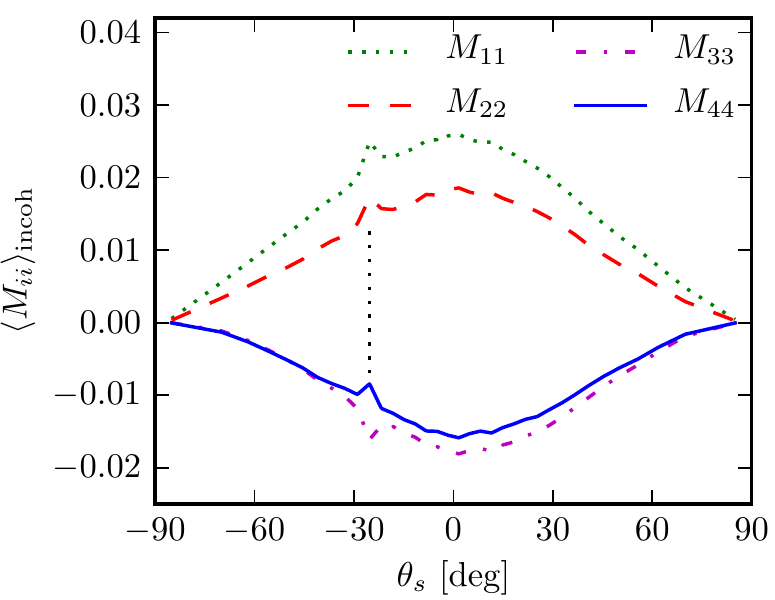}
    \caption{\label{fig:2-cut}
        The incoherent contribution to the diagonal Mueller matrix elements in
        the plane of incidence (parameters as in Fig.~\ref{Fig:2}). The
        vertical dotted line indicates the backscattering direction.
    }
\end{figure}

\begin{figure*}[th]
    \centering
    \includegraphics*[width=1.65\columnwidth]{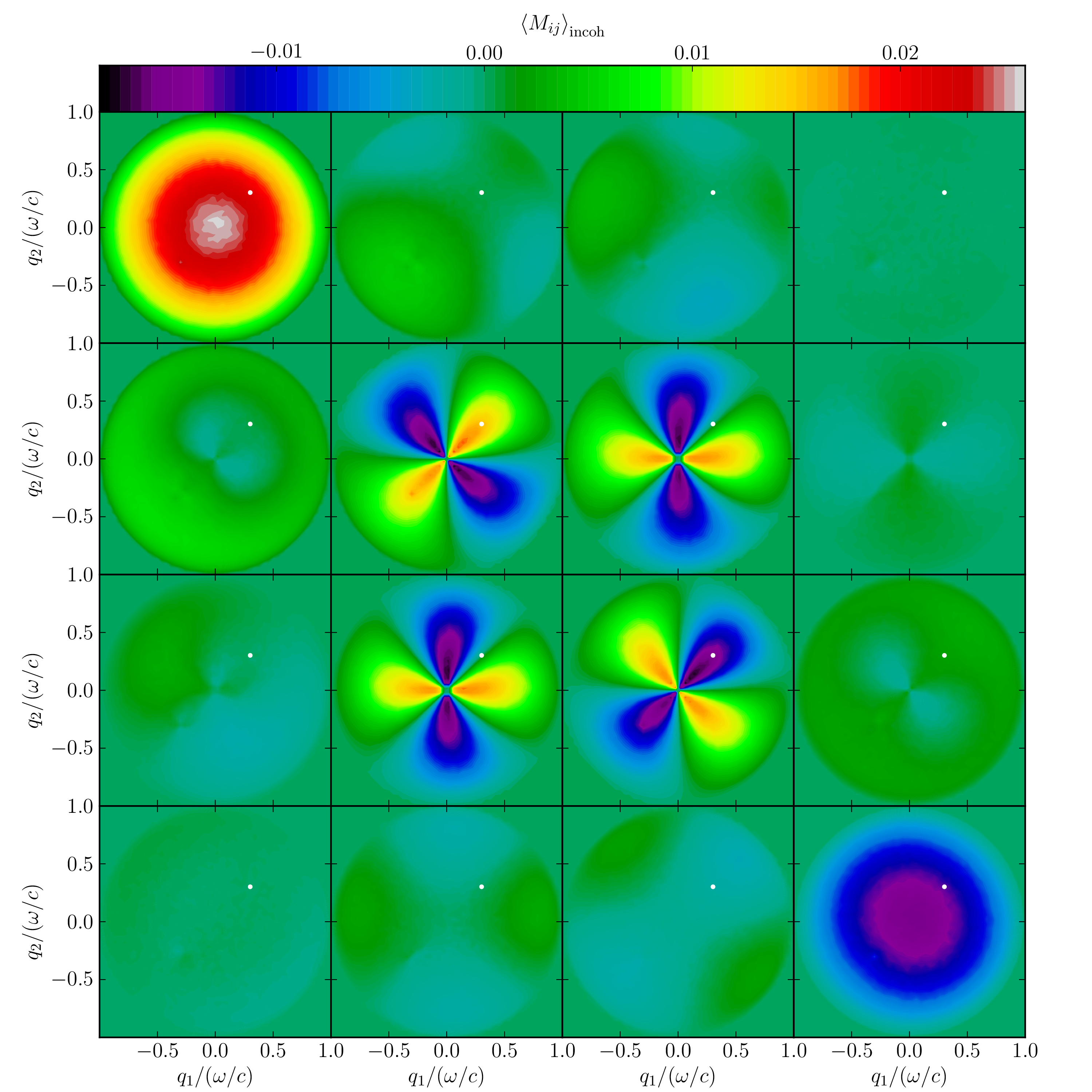}
    \caption{\label{Fig:2}(Color online)
        Same as Fig.~\protect\ref{Fig:1}, but now for angles of
        incidence $(\theta_0,\phi_0)=(25^\circ, 45^\circ)$.}
\end{figure*}

The results presented in Fig.~\ref{Fig:2} were obtained for angles of
incidence $(\theta_0,\phi_0)=(25^\circ, 45^\circ)$, and display some
interesting features. The elements
$\left<M_{11}\right>_{\text{incoh}}$,
$\left<M_{22}\right>_{\text{incoh}}$, and
$\left<M_{33}\right>_{\text{incoh}}$ contain a (weak) enhanced backscattering peak at $\pvec{q}= -\pvec{k}$ (Fig.~\ref{fig:2-cut}). The element
$\left<M_{44}\right>_{\text{incoh}}$ appears to have a dip in the retroreflection
direction. This dip is not present in the results of a calculation
based on small-amplitude perturbation theory to the lowest (second)
order in the surface profile function, and is therefore a multiple
scattering effect, just as the enhanced backscattering peak is.  In
contrast to what was the case for normal incidence, the elements
$\left<M_{31}\right>_{\text{incoh}}$ and
$\left<M_{24}\right>_{\text{incoh}}$ are no longer zero.


If we denote the ensemble average of the contribution to a \emph{normalized} element of the Mueller matrix from the light that has
been scattered incoherently by $m_{ij} =
\left<M_{ij}\right>_{\text{incoh}}/\left<M_{11}\right>_{\text{incoh}}$, we can
estimate the order of magnitude of the Mueller matrix elements by
calculating the quantities $s_{ij}=\left< |m_{ij}(\pvec{q})|
\right>_{\pvec{q}}$, where $\left<f(\pvec{q})\right>_{\pvec{q}}= \int
\!\dint^2q_\parallel \,f(\pvec{q})/ \pi(\omega/c)^2$, and the integral
over $\pvec{q}$ is taken over the circular region
$0<q_\parallel<\omega/c$.  It was found that $s_{11}$, $s_{22}$,
$s_{23}$, $s_{32}$, $s_{33}$, $s_{44}$ are of ${\mathcal O}(1)$;
$s_{12}$, $s_{13}$, $s_{21}$, $s_{34}$, $s_{42}$, $s_{43}$ are of
${\mathcal O}(0.1)$; and $s_{14}$, $s_{24}$, $s_{31}$, $s_{41}$ are of
${\mathcal O}(0.01)$. These results are only weakly dependent on the
polar angle of incidence $\theta_0$, for the values of $\theta_0$
assumed in this study.

In conclusion, we have presented a new approach to the
calculation of all sixteen elements of the Mueller matrix for light
scattered from a two-dimensional, randomly rough, lossy metal
surface, for arbitrary values of the polar and azimuthal angles of
incidence. It is based on a rigorous numerical solution of the reduced
Rayleigh equation for the scattering of p- and s-polarized light
from a two-dimensional rough surface of a penetrable medium, that
captures multiple-scattering processes of all orders.  The results
display multiple scattering effects in certain matrix elements, such
as an enhanced backscattering peak in the retroreflection direction,
or an unexpected dip in the same direction. The matrix elements also
display symmetry properties that, for normal incidence, agree with
those predicted by Bruce~\cite{doi:10.1080/13616679809409826}.

The physical implications of the approach and results of this Letter
point to better understanding of the polarimetric properties of random
surfaces. Such knowledge may be critical for improved photovoltaic and
remote sensing applications and has the potential of being used to
engineer surface structures with well-defined polarization properties
of the light interacting with them.


\begin{acknowledgements}
The authors are grateful for fruitful interactions with M. Lindgren, M.\ Kildemo, I.S.\ Nerb{\o}, and L.M.\ Sandvik Aas.  The research of P.A.L., T.N., and I.S. was supported in part by NTNU by the allocation of computer time. The research of A.A.M. was supported in part by AFRL contract FA9453-08-C-0230.
\end{acknowledgements}

%

\end{document}